\documentclass[10pt]{iopart}
\usepackage{hyperref}
\usepackage{graphicx}
\usepackage{subfig}
\usepackage{color}
\usepackage{ulem}
\usepackage{amssymb}
\usepackage{cite}

\begin{document}
\title[]{Effects of plasma nonuniformity on toroidal Alfv\'en eigenmode nonlinear decay}
\author{Zhiwen Cheng$^1$,   Kexun Shen$^1$,  and Zhiyong Qiu$^{1,2}$}
\address{$^1$ Institute for Fusion Theory and Simulation, School of Physics, Zhejiang University, Hangzhou, China}
\address{$^2$ Center for Nonlinear Plasma Science and C.R. ENEA Frascati, C.P. 65, 00044 Frascati, Italy}
\ead{zqiu@ipp.ac.cn}
\begin{abstract}
    The parametric decay of toroidal Alfv\'en eigenmode (TAE) in nonuniform plasmas is investigated using nonlinear gyrokinetic equation. It is found that, the plasma nonuniformity not only significantly enhances the nonlinear coupling cross-section, but also qualitatively modifies the decay process. Specifically, the condition for spontaneous decay becomes the toroidal mode number of the  sideband TAE being  higher than that of the pump TAE, instead of the frequency of the sideband TAE being lower than the pump TAE in uniform plasmas. The consequences on TAE saturation and energetic particle transport are also discussed.    
\end{abstract}
\noindent{toroidal Alfv\'en eigenmode, nonlinear mode coupling, gyrokinetic theory, parametric decay instability, ion induced scattering}
\maketitle
\ioptwocol
\section{Introduction}

Shear Alfv\'en waves (SAWs) \cite{HAlfvenNature1942} correspond to transverse electromagnetic oscillations  propagating along  equilibrium magnetic field,  with the characteristic Alfv\'en   velocity determined by the magnetic field and plasma density $V_A\equiv B_0/\sqrt{4\pi\rho_m}$. Here, $B_0$ is the equilibrium magnetic field amplitude, $\rho_m=N_0m_i$ is the mass density, $N_0$ is the equilibrium particle density, and $m_i$ is the ion mass. In magnetically confined fusion plasmas, SAWs 
can be resonantly excited by energetic particles (EPs) including fusion alpha particles \cite{YKolesnichenkoVAE1967,AMikhailovskiiSPJ1975,MRosenbluthPRL1975,LChenPoP1994,SPinchesPoP2015,LChenRMP2016}, 
and in turn, induce EP anomalous transport loss across magnetic surfaces, resulting in plasma 
performance degradation and possibly damage of plasma facing components \cite{IPBNF1999,AFasoliNF2007,RDingNF2015}. In tokamak plasmas with nested magnetic surfaces,  due to equilibrium magnetic geometry and plasma nonuniformities,  SAW frequency varies continuously in the  radial direction, and constitutes a continuous spectrum, inside which forbidden gaps form due to symmetry breaking periodic modulation of Alfv\'en velocity  along the magnetic field line. Consequently, 
SAW instabilities can be excited  as various EP continuum modes (EPMs)  \cite{LChenPoP1994} or discrete Alfv\'en eigenmodes (AEs) inside the continuum gaps, e.g. toroidal Alfv\'en eigenmode (TAE)  \cite{CZChengAP1985,LChenVarenna1988,GFuPoFB1989}.  The EP anomalous transport rate is determined by the    amplitude and spectrum of the    electromagnetic perturbations induced by the SAW instabilities  \cite{LChenJGR1999,MFalessiPoP2019}, which highlights the 
significance of investigating the SAW instability  nonlinear dynamics and saturation mechanisms \cite{HBerkPoFB1990c,LChenPoP2013,FZoncaNJP2015,LChenRMP2016}. In the past decades, using the representative TAE as a paradigm case, nonlinear saturation of SAW instabilities are  extensively studied both  experimentally and theoretically  \cite{HBerkPoFB1990c,YTodoPoP1995,JLangPoP2010,SBriguglioPoP2014,JZhuPoP2013,HZhangNF2022,DSpongPoP1994,TSHahmPRL1995,FZoncaPRL1995,LChenPPCF1998,YTodoNF2010,LChenPRL2012,ZQiuPoP2016,ZQiuNF2017,ZQiuPRL2018,ZQiuNF2019a,LChenNF2022,LChenNF2023}.  An important class of  processes for  SAW nonlinear  saturation  is nonlinear mode-mode coupling  describing the mode evolution due to interacting with other collective plasma oscillations \cite{RSagdeevbook1969}, which is expected to be  more  important  in burning plasma of future reactors than present day tokamaks,  as there  could be a rich spectrum of SAW instabilities excited 
simultaneously with comparable linear growth rates  \cite{AFasoliNF2007,LChenRMP2016,TWangPoP2018,ZRenNF2020,MSchnellerPPCF2015,PLauberPR2013,SPinchesPoP2015}. These SAW instabilities can  then 
interact with each other, affecting the  complex spectrum evolution and ultimate confinement of the reactor. 

Among various   nonlinear mode coupling processes, one of the channels expected to crucially affect the  TAEs nonlinear dynamics is  
nonlinear   ion induced scattering \cite{RSagdeevbook1969}, in which a TAE decays into another counter-propagating TAE and a heavily ion Landau damped ion sound quasi-mode (ISM). This process  was originally  investigated by Hahm et al  using drift kinetic theory \cite{TSHahmPRL1995},  based on which  the  governing equations  describing  the   TAE   spectral energy  cascading  are derived and solved, yielding  the saturated spectrum 
and overall electromagnetic perturbation magnitude.  The consequent  bulk ion heating due to ion Landau damping of the ISM has also been investigated, providing a potential ``alpha-channeling" mechanism to effectively transfer energy of fusion alpha particles to fuel ions  \cite{TSHahmPST2015,NFischPRL1992}. While the  analysis  of Ref. \cite{TSHahmPRL1995} was limited to the 
long wavelength with $k_\bot^2\rho_i^2\ll \omega/\Omega_{ci}$ regime where the parallel polarization nonlinearity dominates, 
 this analysis was extended in Ref.   \cite{ZQiuNF2019a} to   the short wavelength kinetic  regime with  $k_\bot^2\rho_i^2\gg \omega/\Omega_{ci}$  \cite{LChenEPL2011} more relevant to  the next generation reactors.  Here,   $k_\bot$ is  the perpendicular wavenumber, $\rho_i$ is  the ion Larmor radius, and $\Omega_{ci}$ is the ion cyclotron frequency. 
At short wavelengths, the dominant nonlinear terms are   Reynolds and Maxwell stresses   in the radially fast varying inertial layer,  and  nonlinear gyrokinetic theory \cite{EFriemanPoF1982} is mandatory to capture the crucial  physics,  as addressed in Ref. \cite{ZQiuRMPP2023}. 
Ref. \cite{ZQiuNF2019a}  yields a significantly    enhanced nonlinear coupling cross-section, and       consequently,  much lower    TAE saturation level and the induced resonant circulating EP transport rate. 

The analysis of both Refs. \cite{TSHahmPRL1995} and \cite{ZQiuNF2019a} assumed uniform thermal plasma condition. Here, for uniform thermal plasma, we mean the thermal ion diamagnetic drift frequency $\omega_{*i}$ is much lower than thermal ion  transit frequency, while other effects, e.g., the thermal plasma nonuniformity contribution to  SAW continuum, are well  preserved in the analysis. If thermal plasma nonuniformity is accounted for, the parametric decay process of TAE into ISM could be qualitatively changed, with crucial effects entering through the diamagnetic drift frequency contribution to  ISM \cite{ZQiuIAEAFEC2023}.  One possibility is, the ISM becomes a drift wave (DW) if the diamagnetic frequency is comparable to the mode frequency,  while the mode is weakly Landau damped. The direct  scattering of TAE by  DW  was investigated in Ref. \cite{LChenNF2022}, where it was shown that TAE instability  can be significantly reduced or even suppressed by ambient DWs.  Another possibility is, the ISM becomes a drift sound wave (DSW) quasi-mode where  ion Landau damping is still significant.   The scattering of TAE by DSW is the main focus of the present work.  

Recently, it is demonstrated in Ref.  \cite{LChenPoP2022} that the parametric decay process of kinetic Alfv\'en waves (KAWs) in nonuniform plasma could be both quantitatively and qualitatively different from that in  uniform plasma. Analyzing a pump KAW decaying into a counter-propagating KAW (``backward scattering") and a DSW quasi-mode, it is found that, the nonlinear coupling is significantly  enhanced   by $\mathcal{O}(\omega_{*i}/\omega_s)$. The  resultant saturated KAW spectrum is asymmetric in $k_y$ (corresponding to $k_{\theta}$ in a toroidally confined plasma), implying,   potentially,  an additional convective  component of radial transport. 
This suggests to investigate   the effects of  plasma nonuniformity on the parametric decay  of TAE, due to the clear  correspondence  between  these two processes.  In particular, the potential modification to TAEs  nonlinear saturation dynamics that  may significantly affect the efficiency of a reactor will be addressed here. 

Thus, in this work, we   generalized the theory of  TAE parametric decay   to  include finite thermal plasma diamagnetic response,  and focused on the  three-wave interaction process  as a starting point for  addressing TAE nonlinear  spectral cascading  and anomalous EP transport. Specifically, the nonlinear dispersion relation for TAE parametric decay in short 
wavelength limit is derived using nonlinear gyrokinetic theory. Based on the analytic results, we then analyze the effects induced 
by plasma nonuniformity, including  enhancement of scattering cross-section, modification of  the dominant scattering process,  and change  of spontaneous 
decay condition. 

The rest of this paper is organized as follows. In Sec.  \ref{sec:model}, the gyrokinetic  theoretical model  is introduced, which is  used  in Sec. \ref{sec:PDI}  to derive  the nonlinear parametric dispersion 
relation. The condition for the nonlinear process to occur is also analyzed, and the implications on TAE saturation are  discussed.  
Finally,   a brief summary is presented in Sec. \ref{sec:summary}.

\section{Theoretical model}\label{sec:model}

\begin{figure}[h]
    \includegraphics[width=0.5\textwidth]{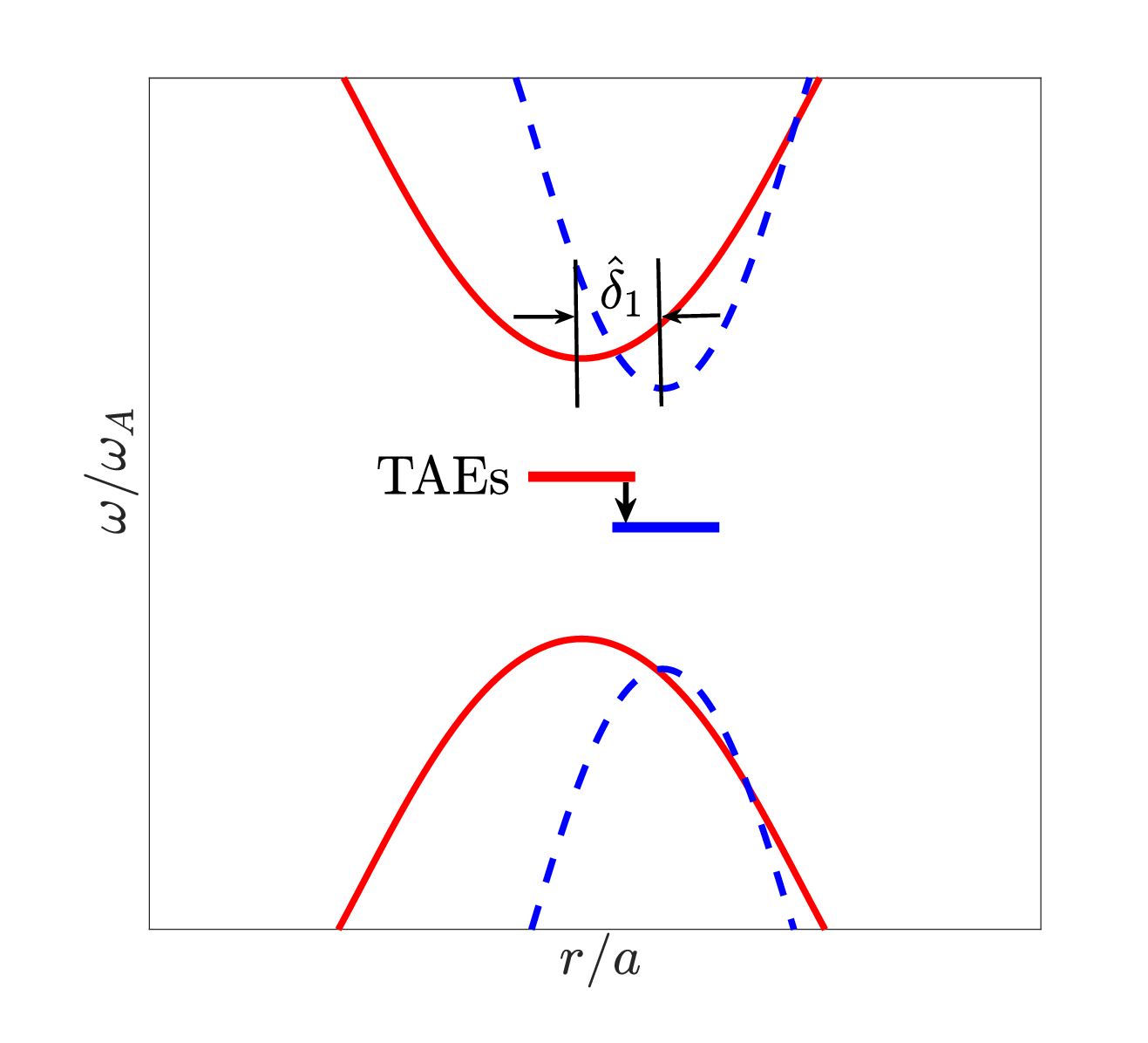}    
    \caption{Cartoons for TAE parametric decay. Here, the horizontal axis is $r/a$ in arbitrary units, the vertical axis is the normalized frequency to $\omega_A=v_A/(qR_0)$ with $q$ the safety factor and $R_0$ the magnetic axis major radius.  The solid and dashed curves represent the SAW continuum with toroidal mode numbers $n_0$ and $n_1$,  and $\hat{\delta}_1$ represents the radial misalignment between  $\Phi_0$ and $\Phi_1$. }
      \label{fig:PDI_cartoon}
\end{figure}

To investigate the effects of plasma nonuniformity on TAE  nonlinear saturation, we consider the process of a pump TAE $\mathbf{\Omega}_0=(\omega_0, \bi{k}_0)$ spontaneously decaying 
into a counter-propagating sideband  toroidal Alfv\'en mode (TAM)  $\mathbf{\Omega}_1=(\omega_1, \bi{k}_1)$ and a low-frequency electrostatic drift sound wave  (DSW) quasi-mode
$\mathbf{\Omega}_s=(\omega_s, \bi{k}_s)$,  which could be  strongly ion Landau damped. Here,  the nonlinearly excited sideband  TAM  could   be a  TAE in the gap or a small scale kinetic TAE (KTAE) \cite{AHasegawaPoF1976,LChenRMPP2021,RMettPoFB1992,FZoncaPoP1996},  depending on the specific parameters, as will be further  addressed  and clarified  in Sec.  \ref{sec:analysis}.   An illustrative plot of TAE parametric decay is given in Fig. \ref{fig:PDI_cartoon}.  Taking  
$\mathbf{\Omega}_0=\mathbf{\Omega}_1+\mathbf{\Omega}_s$ as the matching condition, and   for the low-$\beta$ parameter regime of interest for tokamak plasmas,  introducing the electrostatic potential $\delta\phi$ along with the parallel component of the 
vector potential $\delta A_\parallel$ as field variables, we then have $\delta\phi=\delta\phi_0+\delta\phi_1+\delta\phi_s$, with the subscripts $0, 1$ and $s$ denoting 
$\mathbf{\Omega}_0$, $\mathbf{\Omega}_1$ and $\mathbf{\Omega}_s$, respectively. Furthermore,    $\delta\psi\equiv \omega\delta A_\parallel /(ck_\parallel)$ is also introduced below for convenient treatment of the inductive parallel electric field,  which allows writing the linearized ideal MHD Ohm's law as  
$\delta\psi=\delta\phi$. 
For the parametric decay of a given  finite amplitude pump TAE,  one can include only the bulk plasma nonlinear contribution, while effects of EPs crucial for the TAE excitation is typically negligible for the nonlinear mode coupling process.

For  TAEs with most unstable modes  characterized with    $n\gtrsim \mathcal{O} (10)$ \cite{TWangPoP2018,ZRenNF2020},   the   ballooning-mode representation is adopted  in the $(r,\theta,\phi)$ field-aligned toroidal flux coordinate  \cite{JConnorPRL1978}
\begin{center}
    \begin{eqnarray}
        \delta\phi_0&=&A_0 \rme^{\rmi(n_0\phi-\hat{m}_0\theta-\omega_0 t)}\sum_j \rme^{-\rmi j\theta}\Phi_0(x-j)+c.c.,\nonumber\\
        \delta\phi_1&=&A_1 \rme^{\rmi(n_1\phi-\hat{m}_1\theta-\omega_1 t)}\sum_j \rme^{-\rmi j\theta}\Phi_1(x-j+\delta_1)+c.c..\nonumber
    \end{eqnarray}
\end{center}
Here, $A$ stands for the radial envelope, $n$ is  the toroidal number, $m=\hat{m}+j$ with $\hat{m}$ being  the reference poloidal mode number,  
$x=nq-\hat{m}\simeq nq'(r-r_0)$, $r_0$ is the TAE localization position satisfying $|n_0q(r_0)-\hat{m_0}|\simeq 1/2$, $\Phi$ is  the 
fine radial structure associated with $k_\parallel$ and magnetic shear, $\delta_1\equiv(n_1-n_0)q+\hat{m}_0-\hat{m}_1\mp 1$ is a small normalized radial shift accounting for possible misalignment 
of TAM radial mode structure with respect to that of the pump TAE. 
Meanwhile, the perturbation of DSW $\mathbf{\Omega}_s$ can be expressed as
    \begin{equation}
        \delta\phi_s=A_s \rme^{\rmi(n_s \phi-m_s\theta-\omega_st)}\Phi_s.\nonumber
    \end{equation}
Note that  for DSW, the corresponding typical distance between mode rational surfaces could be  much narrower  than that 
of pump TAE, i.e. $1/|n_sq'(r_s)|\ll 1/|n_0q'(r_0)|$   due to $n_s\gg n_0$.  

Nonlinear gyrokinetic theory is necessary for the crucial physics of ion-induced scattering in the short wavelength regime \cite{LChenEPL2011,LChenRMP2016,ZQiuRMPP2023}. The governing equations describing the 
nonlinear interactions among TAMs and DSW can be derived from the quasi-neutrality condition
\begin{equation}
    \frac{N_0e^2}{T_i}\left(1+\frac{T_i}{T_e}\right)\delta\phi_k=\sum_{j=e,i}\langle qJ_k\delta H_k\rangle_j,
\end{equation}
and  nonlinear gyrokinetic vorticity equation \cite{LChenNF2001}
\begin{eqnarray}
    \fl \frac{c^2}{4\pi\omega_k^2}B\frac{\partial}{\partial l}\frac{k^2_\bot}{B}\frac{\partial}{\partial l}\delta\psi_k+\frac{N_0e^2}{T_i}\left(1-\frac{\omega_{*i}}{\omega}\right)_k
   \left(1-\Gamma_k\right)\delta\phi_k \nonumber\\
    -\sum_{j=e,i}\left \langle qJ_0\frac{\omega_d}{\omega}\delta H\right \rangle_k \nonumber\\
    =-\frac{i}{\omega_k}\sum_{\bi{k=k'+k''}}\Lambda_{k''}^{k'}[\langle e(J_k J_{k'}-J_{k''})\delta L_{k'}\delta H_{k''}\rangle \nonumber\\
    +\frac{c^2}{4\pi}k^{''2}_\bot\frac{\partial_l\delta\psi_{k'}\partial_l\delta\psi_{k''}}{\omega_{k'}\omega_{k''}}].\label{eq:vorticity}
\end{eqnarray}
The  terms on the left hand side  of  Eq. (\ref{eq:vorticity}) stand for field line bending, inertia and curvature-pressure coupling; while the terms on the right hand side are the formally nonlinear terms of   the gyrokinetic Reynolds stress and Maxwell stress due to ion and electron contribution, respectively. 
Here, $q_j$ is  the electric  charge for   particle specie    $``j"$,  
  the angular brackets $\langle ...\rangle$ 
denotes  velocity space integration, $J_k\equiv J_0(k_\bot\rho)$ with $J_0$ being the Bessel function of zero index accounting for finite Larmor radius (FLR) effects, $\rho=v_{\bot}/\Omega_{c}$  is the  Larmor radius with $\Omega_{c}$ being the cyclotron frequency.  
Furthermore,  $l$ is the arc length along the equilibrium magnetic field line, 
$\omega_{*j}=-i(cT/qB_0)_j\hat{\bi{b}}\times\nabla \ln N_j\cdot\nabla$ is  the diamagnetic drift frequency due to plasma  density nonuniformity,  $\omega_d=(v^2_\bot+2v^2_\parallel)/(2\Omega_cR_0)(k_r \sin\theta+k_\theta\cos\theta)$ is  the magnetic drift frequency,  $\Lambda_{k''}^{k'}=(c/B_0)\hat{\bi{b}}\cdot\bi{ k''\times k'}$ accounts for perpendicular scattering with the constraint of frequency and wavevector matching conditions, 
and $\delta L_k\equiv\delta \phi_k-k_\parallel v_\parallel\delta\psi_k/\omega_k$ is the scalar potential in guiding-center moving frame.  For the sake of simplicity, only plasma density nonuniformity contribution is accounted for in $\omega_*$, while temperature nonuniformity effects are systematically neglected.  
The non-adiabatic particle response   $\delta H_k$ is  derived from the  nonlinear gyrokinetic equation \cite{EFriemanPoF1982}:
\begin{eqnarray}
    (-i\omega+v_\parallel\partial_l+i\omega_d)\delta H_{k}=i\frac{q}{m}\left(\omega\partial_E+\frac{m}{T}\omega_*\right)F_MJ_k\delta L_k \nonumber\\
    -\sum_{\bi{k=k'+k''}}\Lambda_{k''}^{k'}J_{k'}\delta L_{k'}\delta H_{k''}.\label{eq:GKE}
\end{eqnarray}
 
\section{Parametric decay instability}\label{sec:PDI}

In this section, we will derive the governing equations describing the nonlinear parametric decay of TAE following the standard procedure \cite{RSagdeevbook1969}. 
Briefly speaking, the  particle responses to the modes  are derived from the nonlinear gyrokinetic equation,  and  substituted into the quasi-neutrality condition and vorticity equation to 
derive  the   nonlinear  sideband equations,  and yield the nonlinear parametric dispersion relation. 
The particle responses can be derived  by separating  the linear from nonlinear   components  by taking $\delta H_k=\delta H_{k,j}^L+\delta H_{k,j}^{NL}$, with the superscripts ``L" and ``NL" denoting the linear and nonlinear responses, respectively.
It is noteworthy  that, in the WKB limit, the present analysis   is  quite similar to  that of Ref. \cite{LChenPoP2022} for parametric decay of KAWs in nonuniform plasma, with differences due to the peculiar features associated 
with toroidal geometry. 

For the low-frequency DSW with  $\omega_s\sim k_{\parallel s}v_{i}\ll k_{\parallel s}v_{e}$ and  
$|\omega_{d,s}|\ll |v_{i}/(qR_0)|$,  the linear particle  response to $\mathbf{\Omega}_s$ can be straightforwardly  derived as
\begin{equation}
    \delta H^L_{s,e}=0,
\end{equation}
\begin{equation}
    \delta H^L_{s,i}=\frac{e}{T_i}\frac{\omega_s -\omega_{*i,s}}{\omega_s-k_{\parallel s}v_\parallel}F_MJ_s\delta\phi_s.
\end{equation}
For high-n TAMs with  $k_{\parallel T}v_{e}\gg \omega_T\gg k_{\parallel T}v_{i}\gg \omega_{d,i},\omega_{d,e}$, the particle response to $\mathbf{\Omega}_s$ can be derived as
\begin{equation}
    \delta H^L_{T,e}=-\frac{e}{T_e}\left(1-\frac{\omega_{*e,T}}{\omega_T}\right)F_M\delta\psi_T,
\end{equation}
\begin{equation}
    \delta H^L_{T,i}=\frac{e}{T_i}\left(1-\frac{\omega_{*i,T}}{\omega_T}\right)F_MJ_T\delta\phi_T,
\end{equation}
The above expression applies to both the pump TAE and the sideband TAM,  so the subscript   $``T"$   can represent  both    TAE and TAM. 
 Note that  the effects of plasma nonuniformity on the TAMs linear dispersion relation are neglected due to the    $|\omega_{*T}/\omega_T| \sim \tau\sqrt{\beta}|\omega_{*i,s}/\omega_s|$  ordering,  while they are kept here for their potential contribution to the nonlinear coupling, which is derived below.

\subsection{Nonlinear  DSW equation}

We  start from the nonlinear DSW equation.  Noting the  $k_{\parallel s}v_{e}\gg \omega_s, \omega_{d,s}$  ordering, the nonlinear electron response to $\mathbf{\Omega}_s$ can be derived as
\begin{equation}
    \delta H_{s,e}^{NL}=i\frac{\Lambda_0^1}{\omega_0}\frac{e}{T_e}F_M\delta\psi_0\delta\psi_{1^*}. \label{eq:isw_i_nl}
\end{equation}
Here $\Lambda_0^1\equiv (c/B_0)\hat{\bi{b}}\cdot\bi{k}_0\times \bi{k}_1$, 
and the superscript $``*"$ denotes complex  conjugate.  In deriving Eq. (\ref{eq:isw_i_nl}),  the linear electron response  to TAM has been used in the nonlinear term of  the    gyrokinetic equation, as well as  $\omega_0\simeq \omega_1$ and $k_{\parallel s}\simeq 2k_{\parallel 0}\simeq -2k_{\parallel 1}$ from the matching condition.  
The nonlinear ion response to $\mathbf{\Omega}_s$ can be derived noting  the $\omega_s\sim k_{\parallel s}v_{i}\gg \omega_{d,s}$ ordering, and one obtains 
\begin{equation}
    \delta H^{NL}_{s,i}=i\frac{\Lambda_0^1}{\omega_0}\frac{e}{T_i}F_MJ_0J_1\frac{k_{\parallel s}v_\parallel-\omega_{*i,s}}{\omega_s-k_{\parallel s}v_\parallel}
    \delta\phi_0\delta\phi_{1^*}.
\end{equation}
Substituting  the   particle responses of $\mathbf{\Omega}_s$   into the quasi-neutrality condition, 
one obtains the nonlinear  equation of DSW 
\begin{equation}
    \varepsilon_{s*}\delta\phi_s=-i\frac{\Lambda_0^1}{\omega_0}\alpha_{s*}\delta\phi_0\delta\phi_{1^*}.\label{eq:dsw_eq}
\end{equation}
Eq. (\ref{eq:dsw_eq}) describes the  modification to DSW due to the coupling between the pump TAE and the  sideband TAM,  with  $\varepsilon_{s*}\equiv 1+\tau+\tau\Gamma_s\xi_sZ(\xi_s)(1-\omega_{*i,s}/\omega_s)$ being  the linear dielectric function of DSW, 
   $\tau\equiv T_e/T_i$, $\Gamma_s\equiv\langle J^2_sF_M/N_0\rangle$, $\xi_s\equiv \omega_s/|k_{\parallel s}v_{i}|$ and $Z(\xi_s)$ is  the well-known 
plasma dispersion function.  Furthermore,  the nonlinear coupling coefficient $\alpha_{s*}$ is defined as
\begin{eqnarray}
\alpha_{s*}\equiv \tau F_1[1+\xi_s Z(\xi_s)(1-\omega_{*i,s}/\omega_s)]+\sigma_{0*}\sigma_{1*}, 
\end{eqnarray}
with $F_1\equiv\langle J_0 J_1J_sF_M/N_0\rangle$,    
$\sigma_{k*}  \equiv  [1+\tau-\tau\Gamma_k (1-\omega_{*i,k}/\omega_k)]/\left(1-\omega_{*e,k}/\omega_k\right)$ being the ratio between $\delta\psi_k$ and $\delta\phi_k$, and $\sigma_{k*}=1$ corresponding to vanishing parallel electric field. 
Note that  TAEs excited by energetic particles  are typically characterized by $k_{\perp}\rho_h\sim O(1)$  with $\rho_h$ being the characteristic EP orbit width \cite{LChenRMP2016,ZRenNF2020},  i.e., $b_k\equiv k_\bot^2\rho_{i}^2\ll 1$, kinetic effects associated with thermal  FLR effects  are usually not important in $\sigma_{k*}$, 
and $\sigma_{k*}\simeq 1$, 
which is a quite useful approximation for certain parameter regimes. Nonetheless, in the current derivation, $\sigma_{k*}$ is systematically kept for the generality of the theory.

\subsection{Nonlinear  TAM sideband equation}

The nonlinear TAM sideband    $\mathbf{\Omega}_1$  equation can be derived following a similar procedure. Noting that $\mathbf{\Omega}_s$  could be a 
heavily ion Landau damped quasi-mode, it is necessary to keep both $\delta H_s^{L}$ and $\delta H_s^{NL}$ while deriving the nonlinear particle response to TAM sideband $\Omega_1$,  since they can be of the same 
order. The resultant nonlinear particle responses to $\mathbf{\Omega}_1$ are 
\begin{equation}
    \delta H^{NL}_{1,e}=  \left(\frac{\Lambda_0^1}{\omega_0}\right)^2\frac{e}{T_e}F_M|\delta\psi_0|^2\delta\psi_1,
\end{equation}
\begin{eqnarray}
    \delta H^{NL}_{1,i}=i\frac{\Lambda_0^1}{\omega_0}\frac{e}{T_i}F_MJ_0J_s\frac{k_{\parallel s}v_\parallel-\omega_{*i,s}}{\omega_s-k_{\parallel s}v_\parallel}
    \delta\phi_0\delta\phi_{s^*} \nonumber\\
    -\left(\frac{\Lambda_0^1}{\omega_0}\right)^2\frac{e}{T_i}F_MJ_0^2J_1
    \frac{k_{\parallel s}v_\parallel-\omega_{*i,s}}{\omega_s-k_{\parallel s}v_\parallel}|\delta\phi_0|^2\delta\phi_1.\label{eq3}
\end{eqnarray}
Substituting linear and nonlinear responses to $\mathbf{\Omega}_1$ into quasi-neutrality condition, one obtains 
\begin{eqnarray}
    \delta\psi_1=\left[\sigma_{1*}-\left(\frac{\Lambda_0^1}{\omega_0}\right)^2\sigma^{NL}_{1*}|\delta\phi_0|^2\right]\delta\phi_1 \nonumber\\
    +i\frac{\Lambda_0^1}{\omega_0}\beta_{1*}\delta\phi_0\delta\phi_{s^*},\label{eq:qn_tam}
\end{eqnarray}
with
$\sigma_{1*}^{NL}\equiv\tau F_2\left[1+\xi_sZ(\xi_s)\left(1- \omega_{*i,s}/\omega_s\right)\right]-\sigma_{0*}^2\sigma_{1*}$,  
$\beta_{1*}\equiv \tau F_1\left[1+\xi_sZ(\xi_s)(1- \omega_{*i,s}/\omega_s)\right]$,  and    $F_2\equiv \langle J_0^2J_1^2F_M/N_0\rangle$.

On the other hand,   the nonlinear gyrokinetic vorticity equation for  $\mathbf{\Omega}_1$ gives 
\begin{eqnarray}
    \frac{(1-\Gamma_1)}{b_1}\left(1-\frac{\omega_{*i,1}}{\omega_1}\right)\delta\phi_1-\left(\frac{\Lambda_0^1}{\omega_0}\right)^2\frac{\kappa^{NL}_{1*}}{b_1}|\delta\phi_0|^2  \delta\phi_1 \nonumber\\
    -\frac{k^2_{\parallel 1}v^2_A}{\omega_1^2}\delta\psi_1=-i\frac{\Lambda_0^1}{\omega_0}\frac{\beta_{2*}}{b_1}\delta\phi_0\delta\phi_{s^*},\label{eq:vorticity_tam}
\end{eqnarray}
with $\kappa_{1*}^{NL}\equiv (F_2-F_1) [1+\xi_sZ(\xi_s)(1- \omega_{*i,s}/\omega_s)]$,  and 
$\beta_{2*}\equiv F_1 [1+\xi_sZ(\xi_s)
     (1- \omega_{*i,s}/\omega_s)]- \varepsilon_{s*}/\tau+ \sigma_{0*}/\tau$.
Note in deriving Eq. (14), the Maxwell stress makes negligible contribution as  $\mathbf{\Omega}_s$ is  predominantly  electrostatic. 
Combining Eqs. (\ref{eq:qn_tam}) with (\ref{eq:vorticity_tam}), one obtains the nonlinear eigenmode equation of $\mathbf{\Omega}_1$
\begin{equation}
    \left[\varepsilon_{A1}+\left(\frac{\Lambda_0^1}{\omega_0}\right)^2\varepsilon_{A1}^{NL}|\delta\phi_0|^2\right]\delta\phi_1=
    -i\frac{\Lambda_0^1}{\omega_0}\alpha_{1*}\delta\phi_0\delta\phi_{s^*}. \label{eq:tam_eq}
\end{equation}
Here, $\varepsilon_{A1}\equiv (1-\Gamma_1)(1-\omega_{*i,1}/\omega_1)/b_1-k^2_{\parallel 1}v^2_A\sigma_{1*}/\omega_1^2$ is 
the  linear dispersion operator   of $\mathbf{\Omega}_1$ in the WKB limit, 
$\varepsilon_{A1}^{NL}\equiv -\kappa_{1*}^{NL}/b_1+k^2_{\parallel 1}v^2_A\sigma_{1*}^{NL}/\omega_1^2$, 
and  $\alpha_{1*}\equiv \beta_{2*}/b_1+k_{\parallel 1}^2v_A^2\beta_{1*}/\omega_1^2$.

\subsection{Parametric dispersion relation}\label{sec:PDR}

Substituting Eq. (\ref{eq:dsw_eq}) into (\ref{eq:tam_eq}), one obtains,  
\begin{eqnarray}
    &&\left[\varepsilon_{A1}+\left(\frac{\Lambda_0^1}{\omega_0}\right)^2\varepsilon_{A1}^{NL}|\delta\phi_0|^2\right]
    \delta\phi_1\nonumber\\
   &=&- \left(\frac{\Lambda_0^1}{\omega_0}\right)^2\frac{\alpha_{1*}\alpha_{s*}}{\varepsilon_{s*}}|\delta\phi_0|^2\delta\phi_1. \label{eq:parametric_DR}
\end{eqnarray}
Equation  (\ref{eq:parametric_DR})  describes the  evolution of the sideband TAM $\mathbf{\Omega}_1$ due to the nonlinear drive by the  pump TAE.  
Multiplying  both sides of Eq. (\ref{eq:parametric_DR}) with $\Phi_{1^*}$,  integrating  it over the radial length $ \Delta r\gg 1/(n_0q')$, 
and noting that  $\varepsilon_{s*}$ varying much slower than $|\Phi_0|^2$ and $|\Phi_1|^2$ in the  radial direction,  one then obtains, the  nonlinear dispersion relation for TAE parametric decay
\begin{equation}
    \left(\hat{\varepsilon}_{A1}-\Delta|A_0|^2-\hat{\chi}_{1}\varepsilon_{s*}|A_0|^2\right)=-\frac{\hat{C}_1}{\varepsilon_{s*}}|A_0|^2. \label{eq:eigen_DR}
\end{equation}
Here,  $\hat{\varepsilon}_{A1}\equiv \int |\Phi_1|^2\varepsilon_{A1}dr$ is the linear eigenmode dispersion relation of  $\mathbf{\Omega}_1$, having chosen the arbitrary normalization condition $\int |\Phi_0|^2 dr=\int |\Phi_1|^2 dr=1$. 
The terms related to  $\Delta$, $\hat{\chi}_1$ and $\hat{C}_1$ represent the nonlinear frequency shift, ion Compton scattering and shielded-ion scattering,  respectively \cite{LChenEPL2011}, 
with the corresponding expressions being 
$ \Delta  \equiv\langle\langle(\Lambda_0^1/\omega_0)^2 [\sigma_{0*}\sigma_{1*}^2-2(F_1/\Gamma_s)
    (\sigma_{0*}\sigma_{1*}- F_1\sigma_s/\Gamma_s)- F_2\sigma_s/\Gamma_s]/(\tau b_1\sigma_{1*})\rangle\rangle$,  
$\hat{\chi}_1 \equiv  \langle   \langle (\Lambda_0^1/\omega_0)^2
    [ F_2/\Gamma_s - (F_1/\Gamma_s)^2]/(\tau b_1\sigma_{1*})  \rangle  \rangle$,  and 
$\hat{C}_1 \equiv   \langle   \langle ( \Lambda_0^1/\omega_0)^2 (\sigma_{0*}\sigma_{1*}-  F_1\sigma_s/\Gamma_s 
    )^2/(\tau b_1\sigma_{1*})  \rangle \rangle$. 
Furthermore,  $\sigma_s\equiv 1+\tau-\tau\Gamma_s$,  and $\langle\langle ...\rangle\rangle\equiv \int (...)|\Phi_0|^2|\Phi_1|^2dr$ accounts for the 
contribution of  mode structure radial overlapping required for nonlinear mode coupling.  

\begin{figure}[h]
    \centering
    \includegraphics[width=0.45\textwidth]{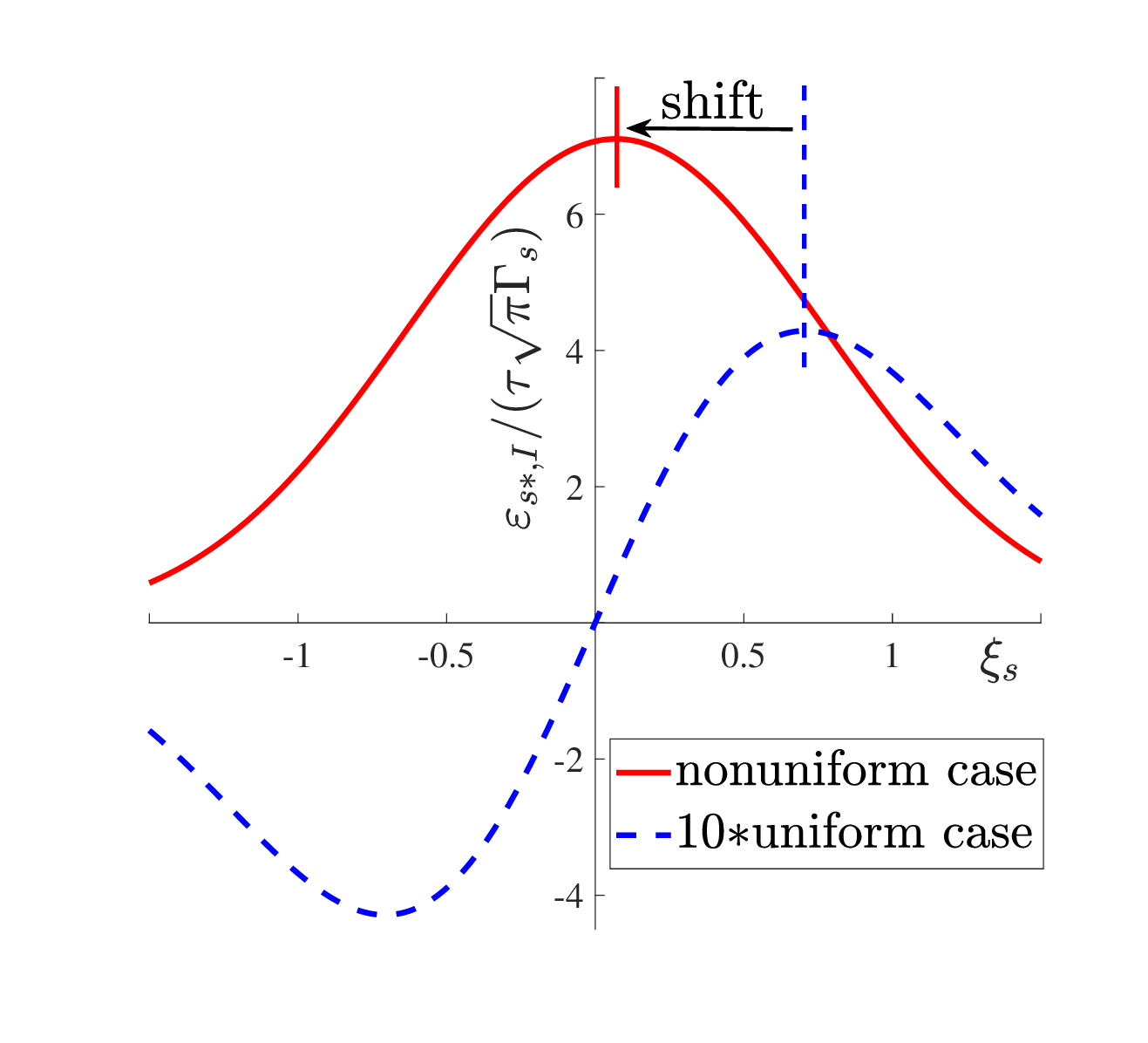}    
    \caption{$\varepsilon_{s*,I}/(\tau\sqrt{\pi}\Gamma_s)$ v.s.  $\xi_s$, with the blue dashed curve standing for uniform case multiplied by   factor 10,  and the red solid curve  for nonuniform case with $qR_0/L_n=10$. }
    \label{fig:xi_s}
\end{figure}

\subsection{Condition for spontaneous decay}\label{sec:analysis}

Equation (\ref{eq:eigen_DR}) is the obtained  parametric dispersion relation, describing the process of a pump TAE  $\mathbf{\Omega}_0$  decaying into a  sideband TAM  $\mathbf{\Omega}_1$  and a  DSW. 
Noting that  DSW is typically  a heavily ion Landau damped quasi-mode with $\varepsilon_{s*,I}$ being comparable to $\varepsilon_{s*,R}$, while the sideband TAM is    a  normal mode of the system, one can expand $\hat{\varepsilon}_{A1}\simeq i(\gamma+\gamma_{A1})\partial_{\omega_1}\hat{\varepsilon}_{A1,R}$, and focus on the stability of the parametric decay process by taking the imaginary part of Eq. (\ref{eq:eigen_DR}),  and   obtains 
\begin{equation}
    \gamma+\gamma_{A1}=\frac{|A_0|^2}{\partial\hat{\varepsilon}_{A1,R}/\partial\omega_{1}}\left(\hat{\chi}_1+\frac{\hat{C}_1}{|\varepsilon_{s*}|^2}\right)\varepsilon_{s*,I}. \label{eq:parametric_gr}
\end{equation}
Here,  $\gamma$ represents  the parametric growth rate,  $\gamma_{A1}\equiv \hat{\varepsilon}_{A1,I}/\partial_{\omega_1}\hat{\varepsilon}_{A1,R}$ is  the linear damping rate of $\mathbf{\Omega}_1$, 
the subscripts $``R"$, $``I"$ denotes real and imaginary parts  respectively, and terms on the right-hand-side stand for the nonlinear drive by ion induced scattering. 
Equation (\ref{eq:parametric_gr}) has the same expression  as the case of uniform plasma \cite{ZQiuNF2019a}, with the crucial  difference from the linear DSW  dispersion relation, 
i.e. the $(1-\omega_{*i,s}/\omega_s)$ term in $\varepsilon_{s*}$  introduced by plasma nonuniformity. 
Equation (\ref{eq:parametric_gr}) is quite complicated, and needs to be investigated in different parameter regimes accounting for the global dispersion relation due to plasma nonuniformity with realistic parameters.  
Some further insights can nonetheless be gained by inspection from Eq. (\ref{eq:parametric_gr}). 

\begin{figure}[h]
    \centering
    \includegraphics[width=0.45\textwidth]{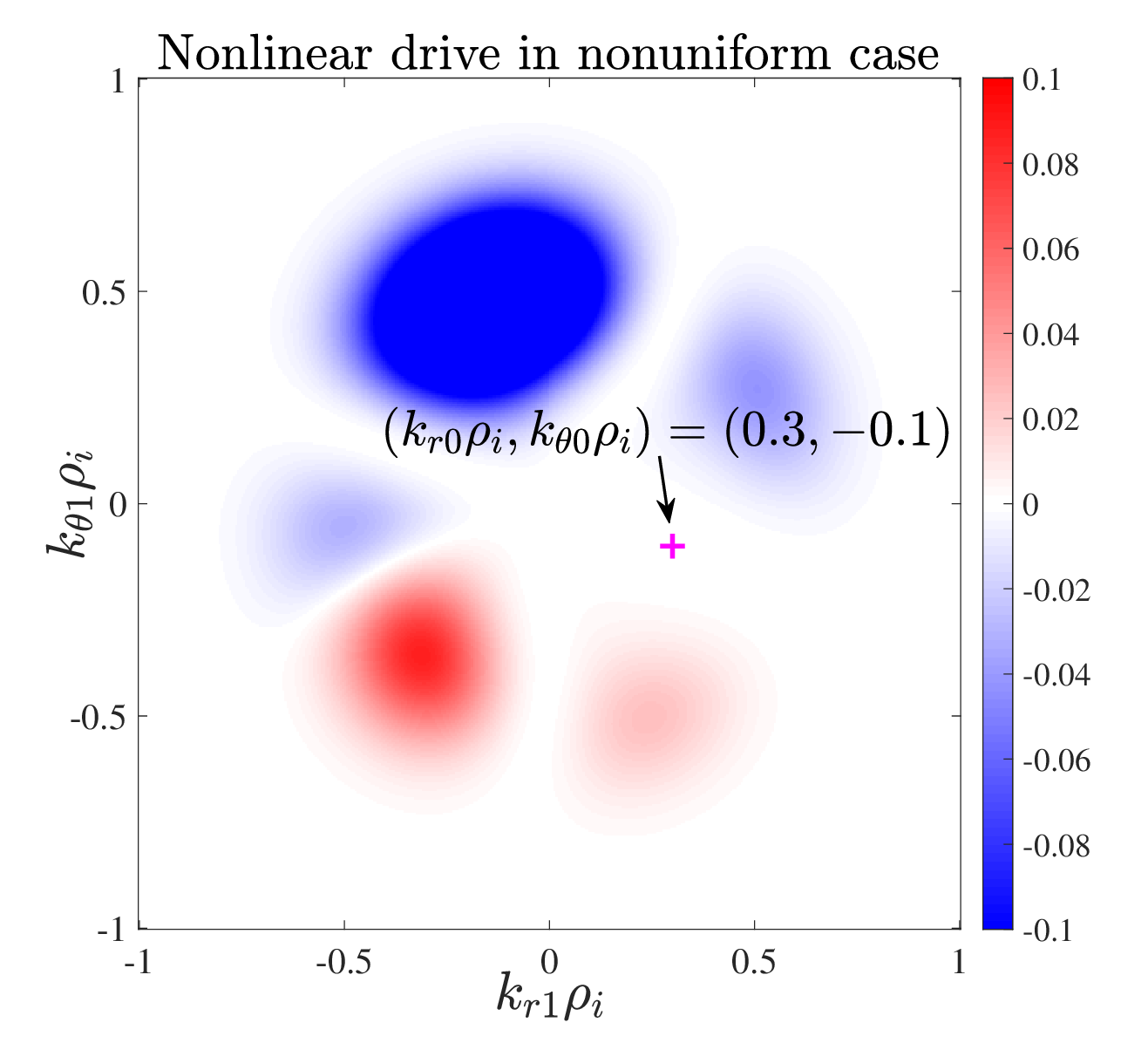}    
    \caption{Contour plot of the nonlinear drive in nonuniform plasma, with the horizontal and vertical axises being $k_{r1}\rho_i$ and $k_{\theta 1}\rho_i$, red and blue colors representing positive and negative. Main parameters used include $qR_0/L_n=10$ and  $(k_{r0}\rho_i,k_{\theta0}\rho_i)=(0.3,-0.1)$. }
    \label{fig:NL_coupling}
\end{figure}

  Noting that  $\varepsilon_{s*,I} \equiv \tau \Gamma_s \xi_s Im(Z_s)(1-\omega_{*i,s}/\omega_s)$, it is expected that the scattering cross-section could  be significantly enhanced  by  $|\omega_{*i,s}/\omega_s|\gg 1$ with respect to the uniform case \cite{ZQiuNF2019a}, as noted in Ref. \cite{LChenPoP2022}.  The $\varepsilon_{s*,I}$ dependence on $\xi_s$ is plotted in Fig. \ref{fig:xi_s}, where the horizontal axis is $\xi_s$, the red solid curve is $\varepsilon_{s*,I}$   with $qR_0/L_n=10$ and $L_n\equiv|N_0/\nabla N_0|$ being the scale length for density nonuniformity, while the blue dashed curve corresponds to the uniform case  with $\omega_{*i}=0$.    Thus the nonlinear coupling is indeed significantly enhanced by   one order of magnitude, as predicted. 
It is also found that, the most effective scattering occurs for $|\xi_s|\ll1$ in the nonuniform plasma cases,   as clearly seen  from  $\varepsilon_{s*,I} \simeq -  \tau \Gamma_s   Im(Z_s) \omega_{*i,s}/|k_{\parallel s} v_i|$ in the $|\omega_{*i,s}/\omega_s|\gg1$ limit, contrary to the well-known $|\xi_s|\sim O(1)$ criterion for most effective ion sound wave Landau damping.  This leads to the conclusion that  the frequency change for each step of decay is small,  suggesting that the cascading process can be approximated as continuous series of vanishing steps.   

\begin{figure}[h]
    \centering
    \includegraphics[width=0.45\textwidth]{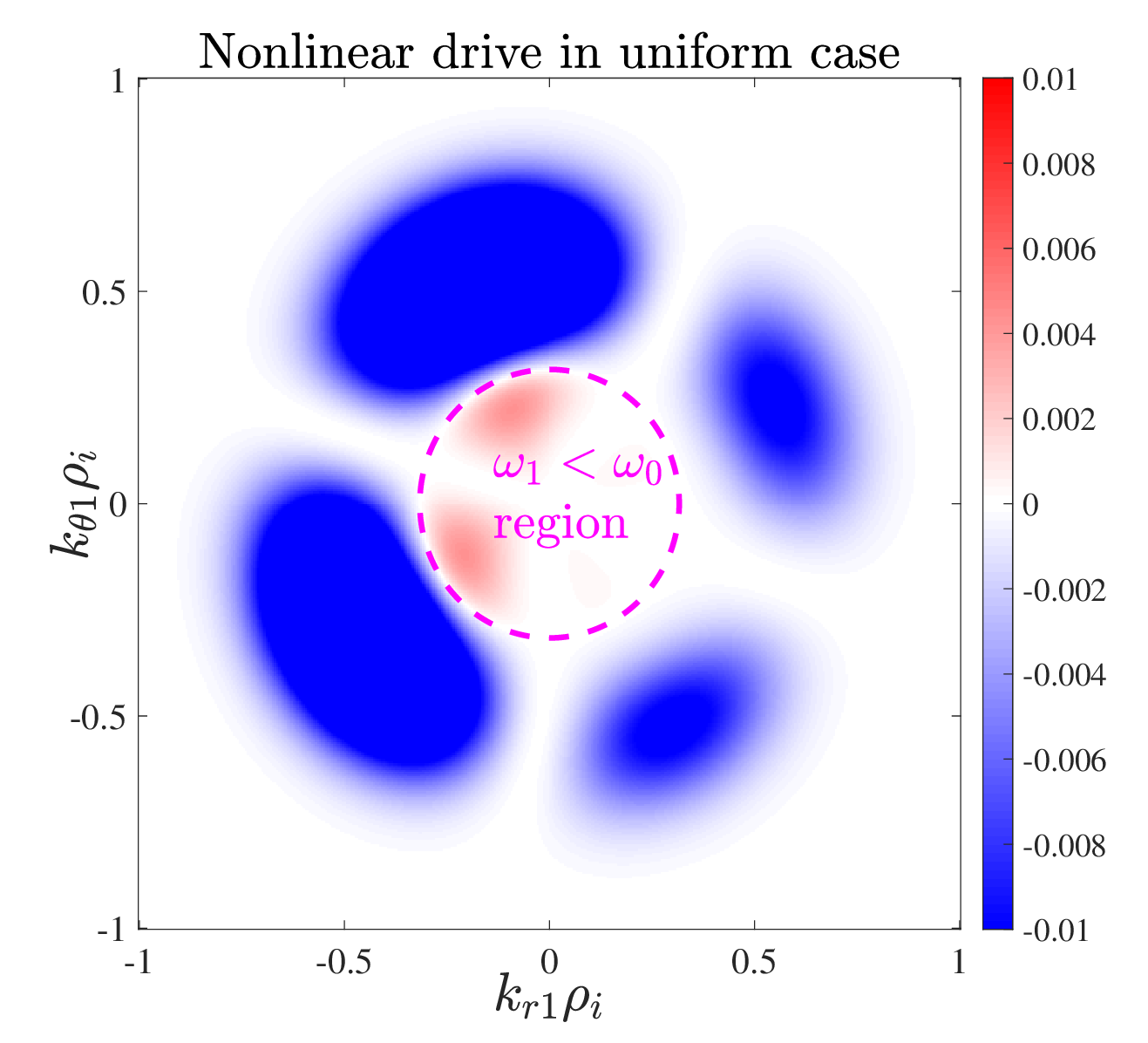}    
    \caption{Contour plot of the nonlinear drive in nonuniform plasmas.  }
    \label{fig:NL_coupling_uni}
\end{figure}

  Noting that $\partial\hat{\varepsilon}_{A1,R}/\partial\omega_{1}>0 $, $\hat{\chi}_1$ is positive definite because of Schwartz inequality, and $\hat{C}_1$ is clearly positive definite from its expression.  The spontaneous decay condition is thus  determined by  the sign of $\varepsilon_{s*,I}$. 
More specifically, the  sign of $\gamma$ is determined by  the sign of $\omega_s-\omega_{*i,s}$ instead of $\omega_s=\omega_0-\omega_1$ in uniform plasma case. 
Thus, while   in uniform plasmas   $\omega_0>\omega_1$ is required for spontaneous decay, i.e.   frequency downward cascading;  however, in  nonuniform plasmas, with $|\omega_{*i,s}|\gg|\omega_s|$, $\omega_{*i,s}\propto k_{\theta1}-k_{\theta0}<0$ is required for    spontaneous decay, 
implying a normal cascading in toroidal mode number $n$, i.e. $|n_1|>|n_0|$.   The nonlinear drive given by the right hand side of Eq. (\ref{eq:parametric_gr}) is shown in Fig. \ref{fig:NL_coupling},  and it is clearly seen that, the condition for spontaneous decay is indeed $n_1>n_0$,  as predicted. 
A crucial qualitative difference with respect to the uniform plasma is that   $\varepsilon_{s*,I}$ can be positive for both positive and negative $\xi_s$,  suggesting downward frequency cascading is no longer needed for spontaneous decay.    For comparison, the nonlinear drive in the uniform plasma limit is plotted in Fig. \ref{fig:NL_coupling_uni}, where the $\omega_1<\omega_0$ criterion for spontaneous decay in uniform plasma is clearly shown \cite{TSHahmPRL1995,ZQiuNF2019a}. 

  It is also found that,     the nonlinear drive is maximized for $k_{\theta1}\lesssim O(1)$,  with predominant contribution from $\varepsilon_{s*,I}$ dependence on $\xi_s$. It is worthy noting that,  in plotting Fig. \ref{fig:NL_coupling}, the WKB dispersion relation of $\Omega_0$ and $\Omega_1$ are used, so the  dependence of $\xi_s$ on $\bi{k}_1$ may have some quantitative discrepancies. However, the general tendency should be the same, due to the $|\xi_s|\ll1$ parameter regime for maximized driven as shown in Fig. \ref{fig:xi_s}.  This suggests the decay into TAE inside the gap  is preferred, while KAW effects are not  crucial.

 The nonlinear scattering is maximized as $\omega_s$ and $\omega_{*i,s}$ have opposite signs, i.e., the DSW rotates in electron diamagnetic drift direction \cite{LChenPoP2022}. This may lead to spectrum asymmetry in $k_{\theta}$,  implying a potential convective component in radial transport, in addition to the usual diffusive component \cite{LChenPoP2022}.  However, this effect is expected to be very small, as the maximized coupling occurs for $|\xi_s|\ll1$, which can also be seen clearly from the asymmetry of the red curve of Fig. \ref{fig:xi_s} with respect to the vertical axis.  Furthermore,  $\hat{\chi}_1$ and $\hat{C}_1$ are typically  comparable to each other, while  the plasma nonuniformity   can lead to  $|\varepsilon_{s*}|\sim |\omega_{*i,s}/\omega_s|\gg 1$. Thus,    the shielded-ion scattering ($\hat{C}_1/|\varepsilon_{s*}|^2$)  can be significantly reduced, and  ion Compton scattering will play the crucial role in saturating the pump TAE.

Equation (\ref{eq:parametric_gr}) can be further simplified in the $b\ll1$ limit.  Noting that,  $\partial\varepsilon_{A1}/\partial\omega_1\simeq 2/\omega_1$, $\hat{\chi}_1\sim c^2 \tau k^2_rk^2_{\theta} b/(\omega^2_0B^2_0)$, $\delta\phi_0\sim \omega_0\delta B_r/(ck_{\parallel 0}k_{\theta0})$, and $\varepsilon_{s*,I} =\sqrt{\pi} \tau\Gamma_s(|\omega_{*i,s}/k_{\parallel s}v_i|) \exp(-\xi^2_s)\simeq \sqrt{\pi}\tau(|\omega_{*i,s}/k_{\parallel s}v_i|)$ since $|\xi_s|\ll1$ for maximized scattering cross-section, one obtains, 
\begin{eqnarray}
\frac{\gamma+\gamma_{A1}}{\omega_1}\simeq \frac{\sqrt{\pi}}{2}\frac{k^2_r}{k^2_{\parallel1}} b_1 \left|\frac{\omega_{*i,s}}{k_{\parallel s}v_i}\right| \left|\frac{\delta B_r}{B_0}\right|^2,
\end{eqnarray}
which yields, 
\begin{eqnarray}
\frac{\gamma+\gamma_{A1}}{\omega_1}\sim O(10^{-2}-10^{-1}).\label{eq:gamma_NL}
\end{eqnarray}
In obtaining Eq. (\ref{eq:gamma_NL}), typical parameters are used, i.e., $b_1\lesssim O(1)$ from the above analysis and Fig. \ref{fig:NL_coupling}, $k^2_r/k^2_{\parallel}\sim O(10^{4}- 10^{5})$, $|\omega_{*i,s}/k_{\parallel s}v_i|\sim qR_0/L_n\sim O(10)$,  and $|\delta B_{r0}/B_0|^2\sim O(10^{-7})$ for typical experimentally observed magnetic fluctuations.  This suggests that, the nonlinear parametric decay process can be important  due to the enhanced scattering by plasma nonuniformity with   $qR_0/L_n\sim 10$. The consequences on TAE final saturation and   plasma heating  will be addressed in a separate publication.

\section{Summary}\label{sec:summary}

In this work,   the parametric decay of toroidal Alfv\'en eigenmode (TAE)  in nonuniform plasmas is investigated using nonlinear gyrokinetic theory,   where a pump TAE decays into a counter-propagating sideband TAE and a low frequency drift sound wave (DSW) quasi-mode.  It is found that the plasma nonuniformity not only significantly enhances the nonlinear  scattering cross-section, but also qualitatively modifies the nonlinear process and the saturated spectrum. Specifically, the main findings are summerized as follows. 
\begin{enumerate}
\item   The nonlinear coupling coefficient is enhanced by $O(\omega_{*i,s}/\omega_s)\sim O(10)$, for typical parameters with $qR_0/L_n\sim O(10)$.  
\item  The condition for spontaneous decay changes from  $\omega_0>\omega_1$ to $|n_1|>|n_0|$, i.e., the process will lead to TAE spectral normal  cascading in toroidal mode number $n$, instead of the downward spectral energy transfer in frequency.  Here, we recall that the subscript ``0" and ``1" denote pump and sideband TAE, respectively.
\item The maximized parametric decay occurs for $|\omega_s|\ll |k_{\parallel s}v_i|$, instead of the typical $|\omega_s|\simeq  |k_{\parallel s}v_i|$ regime in uniform plasmas, due to the modification of the nonlinear term by plasma nonuniformity.  As a consequence, the sideband is expected to be a TAE located inside the toroidicity induced SAW continuum gap, with $|k_{\perp}\rho_i|\lesssim 1$.
\item For typical plasma parameters, this  nonlinear process can strongly occur with $\gamma/\omega_r\sim O(10^{-2}-10^{-1})$ due to the enhanced scattering by plasma nonuniformity, and is expected to significantly  contribute to TAE saturation in burning plasma conditions. 
\end{enumerate}

The  TAE parametric decay process investigated in this work, is a starting point for the analysis of TAE cascading   and final saturation, required for assessing the energetic particle confinement of future reactor burning plasmas. This can be done by  summing up all the  background TAEs that may contribute to Eq. (\ref{eq:tam_eq}), taking the anti-Hermitian part, and taking the continuum limit in frequency, noting the small frequency transfer for maximized scattering.   The obtained equation describes the  TAE spectral energy transfer,  which can be solved for the saturated spectrum and overall electromagnetic 
fluctuations, and finally the consequent  EP anomalous transport rate. These analyses are beyond the scope of the present work, and will be reported in a future publication.

\section*{Acknowledgement}
The authors acknowledge the fruitful discussion with Profs. Liu Chen (Zhejiang University, China and University of California, Irvine)  and Fulvio Zonca (Center for Nonlinear Plasma Science and C.R. ENEA Frascati, Italy and Zhejiang University, China). This work was  supported by the National Key Research and Development  Program of China under Grant No. 2019YFE03020003,   the National Science Foundation of China under Grant Nos. 12275236 and 12261131622, and  Italian Ministry for Foreign Affairs and International Cooperation Project under Grant No. CN23GR02.

\section*{Reference}
\bibliographystyle{iopart-num}
\bibliography{ZQiubib.bib}

\end{document}